\documentclass[pra,groupedaddress,twocolumn]{revtex4-1}

\usepackage{amsmath}    
\usepackage{graphicx}   
\usepackage[raggedright,tight]{subfigure}  
\usepackage[colorlinks=true,citecolor=black,linkcolor=black,backref=true]{hyperref}   

\usepackage{amsfonts}
\usepackage{amssymb}

\usepackage{natbib}
\newcommand{\bra}[1]{\ensuremath{\langle#1|}}
\newcommand{\ket}[1]{\ensuremath{|#1\rangle}}

\newcommand{\bpm}{\begin{pmatrix}}
\newcommand{\epm}{\end{pmatrix}}

\newcommand{\I}{\text{i}}

\newcommand{\MTR}{\text{\sf Tr}}

\newcommand{\dd}[1]{\text{d}#1}

\newcommand{\tum}[1]{\,\text{d}\mu(#1)}
\newcommand{\pp}{p}

\newcommand{\degree}{\ensuremath{^\circ}}

\begin{document}

\title{Generation of maximally entangled states with sub-luminal Lorentz boosts}
\author{Veiko Palge}
\author{Jacob Dunningham}
\affiliation{School of Physics and Astronomy, University of Leeds, Leeds LS2 9JT, United Kingdom}

\begin{abstract}
Recent work \cite{dunningham_entanglement_2009} has studied entanglement between the spin and momentum components of a single spin-$1/2$ particle and showed that maximal entanglement is obtained only when boosts approach the speed of light. Here we extend the boost scenario to general geometries and show that, intriguingly, maximal entanglement can be achieved with boosts less than the speed of light. Boosts approaching the speed of light may even decrease entanglement. We also provide a geometric explanation for this behavior.  
\end{abstract}

\maketitle

\emph{Introduction.}---%
Quantum entanglement is widely held to be the crucial feature that discriminates between quantum and classical physics; it is also at the heart of quantum information theory. While most of the theory of entanglement is non-relativistic, a complete account of entanglement requires that we understand its behavior in the relativistic regime.

Studies in relativistic quantum information have found that single and two particle entanglement becomes an observer dependent phenomenon when viewed from different Lorentz boosted frames \cite{peres_quantum_2002,gingrich_quantum_2002,alsing_entanglement_2002,li_relativistic_2003,peres_quantum_2004,caban_lorentz-covariant_2005,caban_einstein-podolsky-rosen_2006-1,jordan_lorentz_2007,friis_relativistic_2010}. Recent work \cite{dunningham_entanglement_2009} has also investigated 
entanglement between the spin and momentum components of a single particle
and showed that it reaches a maximum value only when boosts approach the speed of light. 
In this paper, however, we demonstrate that maximal entanglement can be obtained for realistic quantum states with boosts less than the speed of light. 
We furthermore show that this behavior can be given a natural geometric explanation.

\begin{figure}[htb]
\includegraphics[width=0.4\textwidth]{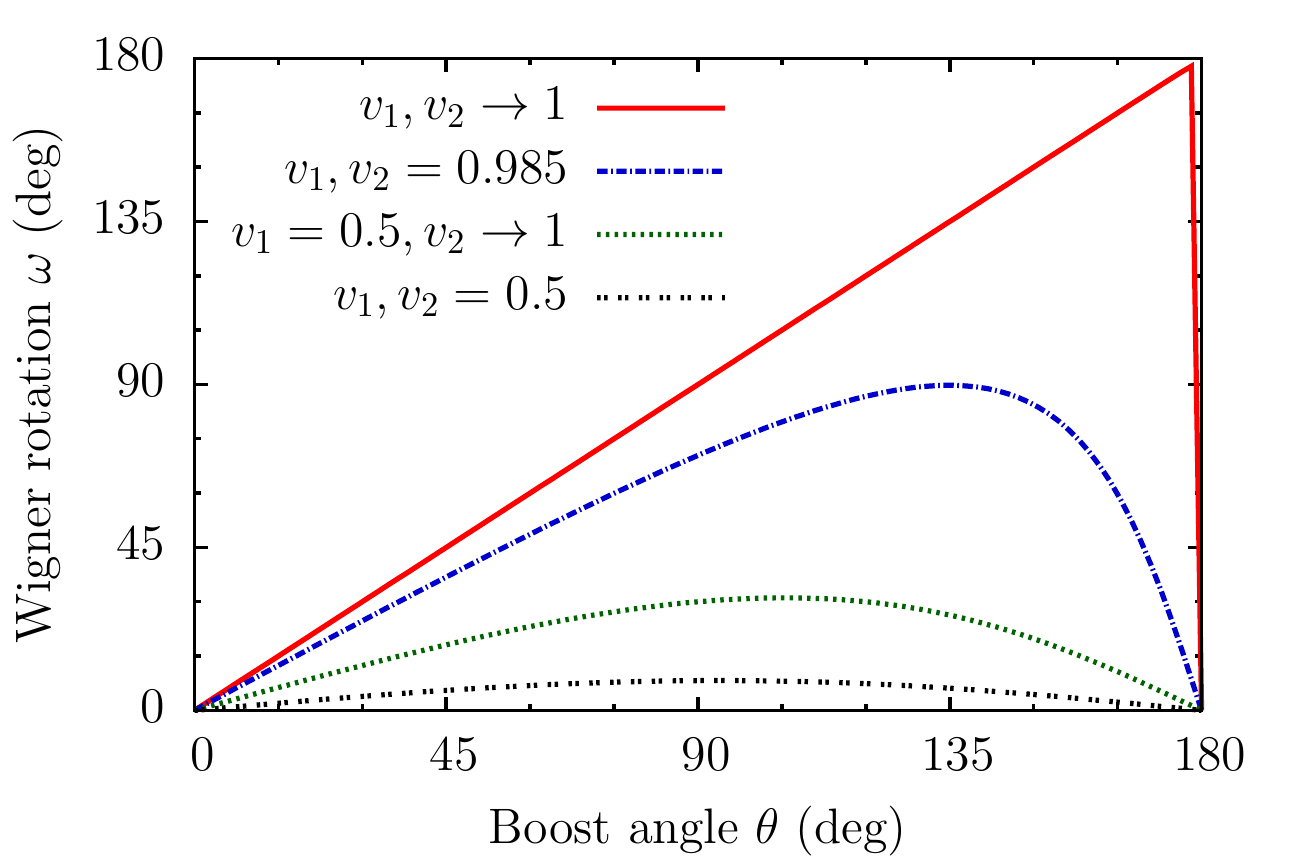}
\caption{\label{fig:fig_1_wigner_rotation}(Color online) Dependence of Wigner rotation on the angle $\theta$ between two boosts.}
\end{figure}

\emph{Properties of Wigner rotation.}---%
We start by reviewing some of the properties of Wigner rotation that are key to our analysis. Wigner rotation arises from the fact that the subset of Lorentz boosts does not form a subgroup of the Lorentz group. Consider three inertial observers $O$, $O'$ and $O''$ where $O'$ has velocity $v_1$ relative to $O$ and $O''$ has $v_2$ relative to $O'$. Then the combination of two canonical boosts $\Lambda(v_1)$ and $\Lambda(v_2)$ that relates $O$ to $O''$ is in general a boost \emph{and} a rotation,
\begin{align}
\Lambda(v_2) \Lambda(v_1) = R(\omega) \Lambda(v_3) \;,
\end{align}
where $R(\omega)$ is the Wigner rotation with 
angle $\omega$. To an observer $O$, the frame of $O''$ appears to be rotated by $\omega$. We will immediately specialize to massive systems, then 
$R(\omega) \in \mathrm{SO(3)}$ and $\omega$ is given by \citep{rhodes_relativistic_2004,halpern_special_1968},
\begin{align}\label{eq:TWRHalfTanFormula}
\tan \frac{\omega}{2} = \frac{\sin \theta}{\cos \theta + D} \;,
\end{align}
where $\theta$ is the angle between two boosts or, equivalently,
$v_1$ and $v_2$, and
\begin{align}\label{eq:theDFactor}
D = \sqrt{\left( \frac{\gamma_1 + 1}{\gamma_1 - 1} \right) \left( \frac{\gamma_2 + 1}{\gamma_2 - 1} \right)} \;,
\end{align}
with 
$\gamma_{1,2} = (1 - v_{1,2}^2)^{-1/2}$. We assume natural units throughout, $\hbar = c = 1$. The axis of rotation specified by $\hat{n} = \hat{v}_2 \times \hat{v}_1$ is orthogonal to the plane defined by $v_1$ and $v_2$. The dependence of Wigner rotation on the angle between two boosts is shown in FIG.~\ref{fig:fig_1_wigner_rotation}.

Several interesting characteristics are immediately noticeable. 
First, for any two boosts with velocities $v_1, v_2$ at an angle $\theta$, the Wigner rotation increases with both $v_1, v_2$, approaching the maximum value $180\degree$ as $v_1, v_2$ approach the speed of light. Second, the maximum value of $\omega$ is bounded by the smaller boost. If $v_1 = 0.5$, then even if $v_2$ becomes arbitrarily close to the speed of light, $\omega$ will be considerably lower than in the case when both boosts approach the speed of light. 
Third, the angle $\theta$ at which the maximum Wigner rotation occurs depends on the magnitudes of both $v_1$ and $v_2$. 
It is worth noting that $\omega$ approaches the maximum value $180\degree$ when both boosts are almost opposite and both $v_1, v_2 \rightarrow 1$. At lower velocities, maximum rotation occurs earlier. We will see below that all these features play an important role in explaining the behavior of entanglement in boosted frames.

\emph{Lorentz boosted single spin-$1/2$ particle.}---%
We will focus on a single massive spin-$1/2$ particle and ask, ``Assuming that spin and momentum are initially in a product state, will they become entangled after two non-collinear Lorentz boosts?'' Consider an observer $O$ who sees the particle in motion with constant momentum. Using basis vectors of the form $\ket{\pp}\ket{\lambda}$, where $\pp$ labels momentum and $\lambda = \pm \frac{1}{2}$ is spin, we can write a generic pure state of the particle as
\begin{align}
\ket{\psi} = \sum_\lambda \int \psi_\lambda(\pp) \ket{\pp}\ket{\lambda} \tum{\pp} \;,
\end{align}
with $\tum{\pp} = (2E(\pp))^{-1}\dd{\pp}$ being the Lorentz invariant integration measure
and the wave function satisfying 
\begin{align}\label{eq:relativisticNormalization}
\sum_\lambda \int |\psi_\lambda(\pp)|^2 \tum{\pp} = 1\;.
\end{align}
To an observer $O''$ who is Lorentz boosted relative to $O$ by $\Lambda^{-1}$ the state of the particle \ket{\psi} appears transformed by $U(\Lambda)$. The action of $U(\Lambda)$ on a basis vector is given by
\begin{align}
U(\Lambda) \ket{\pp}\ket{\lambda} = \sum_{\kappa} \ket{\Lambda\pp}\ket{\kappa} U_{\kappa \lambda}(R(\Lambda, \pp)) \;,
\end{align}
where $U(R) \in \mathrm{SU}(2)$ is the spin-$1/2$ representation of the Wigner rotation $R$. This means that to the observer $O''$ the boosted spin appears rotated by $U(R)$. $U(\Lambda)$ induces the following transformation of the wave function,
\begin{align}\label{eq:unitaryActionOnFiber}
\psi_\lambda (\pp) \mapsto \psi'_{\lambda}(\pp) = \sum_\kappa U_{\lambda \kappa}(R(\Lambda, \Lambda^{-1}\pp)) \psi_\kappa(\Lambda^{-1}\pp) \;.
\end{align}
Since we are interested in knowing the spin state $\rho_S$ according to $O''$, we trace out the momentum degrees of freedom, 
\begin{align}\label{eq:GeneralSpinStateMatrixFormAfterLorentzBoost}
\rho_S &= \mathsf{Tr}_{\pp} \left( U(\Lambda) \ket{\psi} \bra{\psi} U^{\dagger}(\Lambda) \right) \nonumber \\
&= \sum_{\lambda \kappa} \int \psi'_{\lambda}( \Lambda^{-1} \pp ) \psi'^{*}_{\kappa} ( \Lambda^{-1} \pp ) \ket{\lambda} \bra{\kappa} \tum{\pp} \;.
\end{align}
Finally, to quantify how much the entanglement has changed between the spin and momentum degrees of freedom, we calculate the von Neumann entropy of the spin state
\begin{align}
S(\rho_S) = -\MTR(\rho_S \log \rho_S) \;. 
\end{align}
For all boost scenarios to be discussed below we will assume that the particle is boosted in the positive $z$-direction in the canonical way. Writing particle's momentum in Cartesian coordinates, $\pp = (p_x, p_y, p_z)$ and $v_1 = |\pp| / E(\pp)$, the unitary representation of the Wigner rotation takes the form \citep{halpern_special_1968}
\begin{align}\label{eq:HalpernSpinHalfBoostZDirection}
U(R(\Lambda(\xi), \pp)) = 
\bpm	
\alpha  &   \beta (p_x - \I p_y) \\
-\beta (p_x + \I p_y)  &   \alpha 
\epm \;,
\end{align}
with 
\begin{align}
\alpha &= \sqrt{\frac{E + m}{E' + m}} \left( \cosh \frac{\xi}{2} + \frac{p_z}{E + m} \sinh \frac{\xi}{2} \right) \;,
\nonumber \\
\beta &= \frac{1}{\sqrt{(E + m) (E' + m)}} \sinh \frac{\xi}{2} \;,
\end{align}
where $\xi = \mathrm{arctanh}\, v_2$ is the rapidity of the boost in the $z$-direction,
and
\begin{align}
E' = E \cosh \xi + p_z \sinh \xi \;.
\end{align}

\emph{Particle in different boost scenarios.}---%
In a previous paper \cite{dunningham_entanglement_2009}, we studied a single spin-$1/2$ particle in a superposition of two momentum 
delta states 
\begin{align}\label{eq:velocityDeltaState}
\ket{\psi} = \frac{1}{\sqrt{2}} \left( \ket{-p_1} + \ket{p_1} \right) \ket{0} \;,
\end{align}
and showed that boosting the particle induces non-trivial changes of entanglement. A particle whose state is a product of spin and momentum for an observer $O$, appears entangled to a relativistically boosted observer $O''$.

While this provides key insight into relativistic entanglement, it represents a special case and in this paper we will generalize the treatment in two ways. First, because a quantum particle has in general no definite momentum, we will now assume that while the state in the rest frame is a product of spin and momentum as before,
\begin{align}\label{eq:zSpinUpGaussianMomentum}
\ket{\psi} = \int \psi_0(\pp, \pp_0) \ket{\pp}\ket{0} \tum{\pp} \;,
\end{align}
the momentum is given by a superposition of Gaussian wavepackets of finite width $\sigma$
\begin{align}\label{eq:XSymmetricGaussian}
\psi_0 (\pp, \pp_0) = 
\bigg\{&
\frac{1}{N(\sigma)} 
\exp\bigg( -\frac{p_y^2 + (p_z - p_{z0})^2}{2\sigma^2}\bigg) \nonumber \\
&\times
\bigg[
\exp \bigg( -\frac{(p_x - p_{x0})^2}{2\sigma^2} \bigg) \; + \nonumber\\
&\phantom{aa}\times \exp \bigg( -\frac{(p_x + p_{x0})^2}{2\sigma^2} \bigg) 
\bigg]
\bigg\}^{\frac{1}{2}} \;,
\end{align}
where $N(\sigma)$ is normalization. The peaks of Gaussians are symmetrically located at $p_{x0}$ and $-p_{x0}$ from the origin in the momentum space (henceforth $x$-symmetric Gaussian). Second, previously we assumed that the two boosts are orthogonal. 
However, as is evident in FIG.~\ref{fig:fig_1_wigner_rotation}, the geometry of the Wigner rotation is much richer. Boosts at smaller angles tend to result in less Wigner rotation, while larger boost angles produce larger rotation angles; and the magnitude of either boost plays a role as well. 
This suggests that the behavior of entanglement also depends on whether the particle is moving in the same direction as the observer, or in the opposite direction. In order to study this dependency, we will consider three different boost scenarios as follows.

\begin{widetext}
\begin{figure*}[th] 
\subfigure[(Color online) Spin entropy for three boost geometries with different $\theta_i$, all $v_1 = 0.985$.]%
{%
\includegraphics[width=0.35\textwidth]{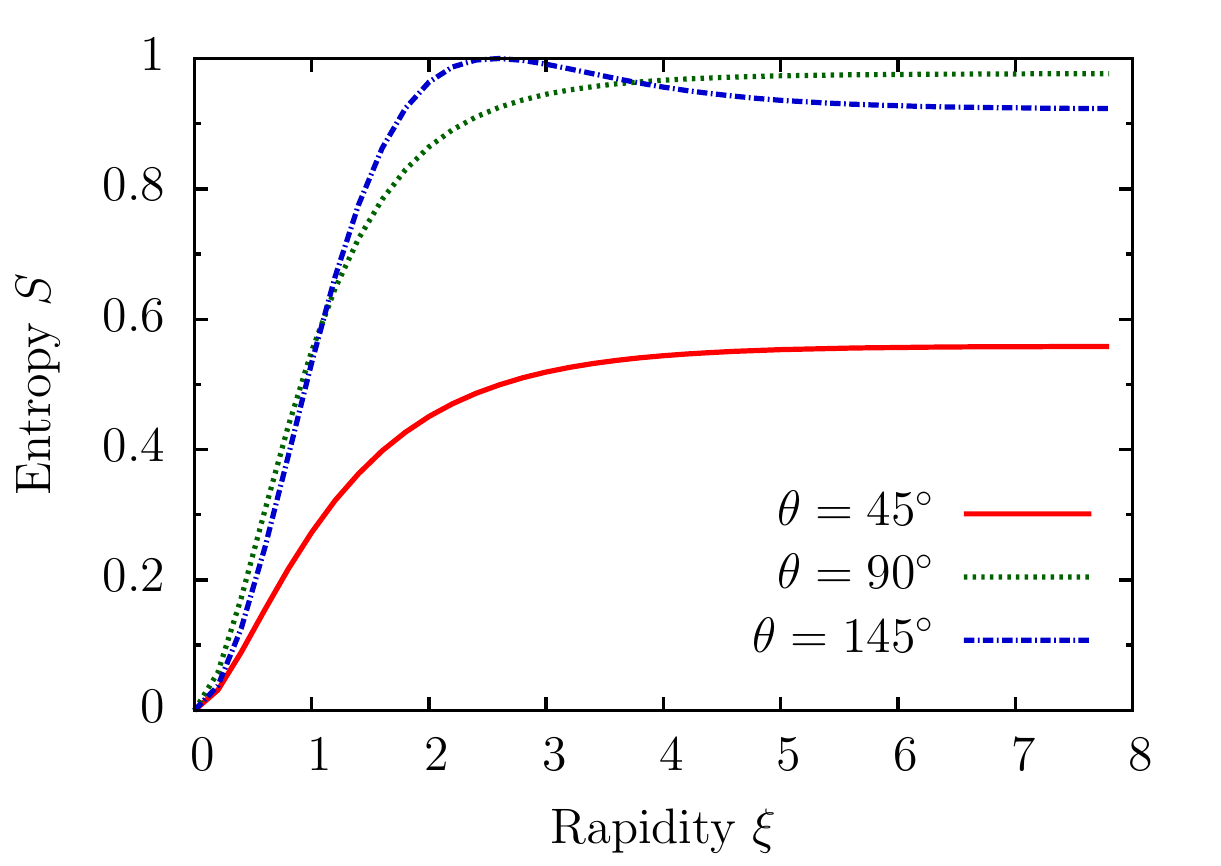}
\label{fig:fig_2a_three_boosts}}%
\hspace{0.5em}
\subfigure[(Color online) Spin entropy for two boost geometries $\theta_{e}, v_1 = 0.999$ and $\theta_{f}, v_1 = 0.99995$, with $\theta > 90\degree$.]{%
\includegraphics[width=0.35\textwidth]{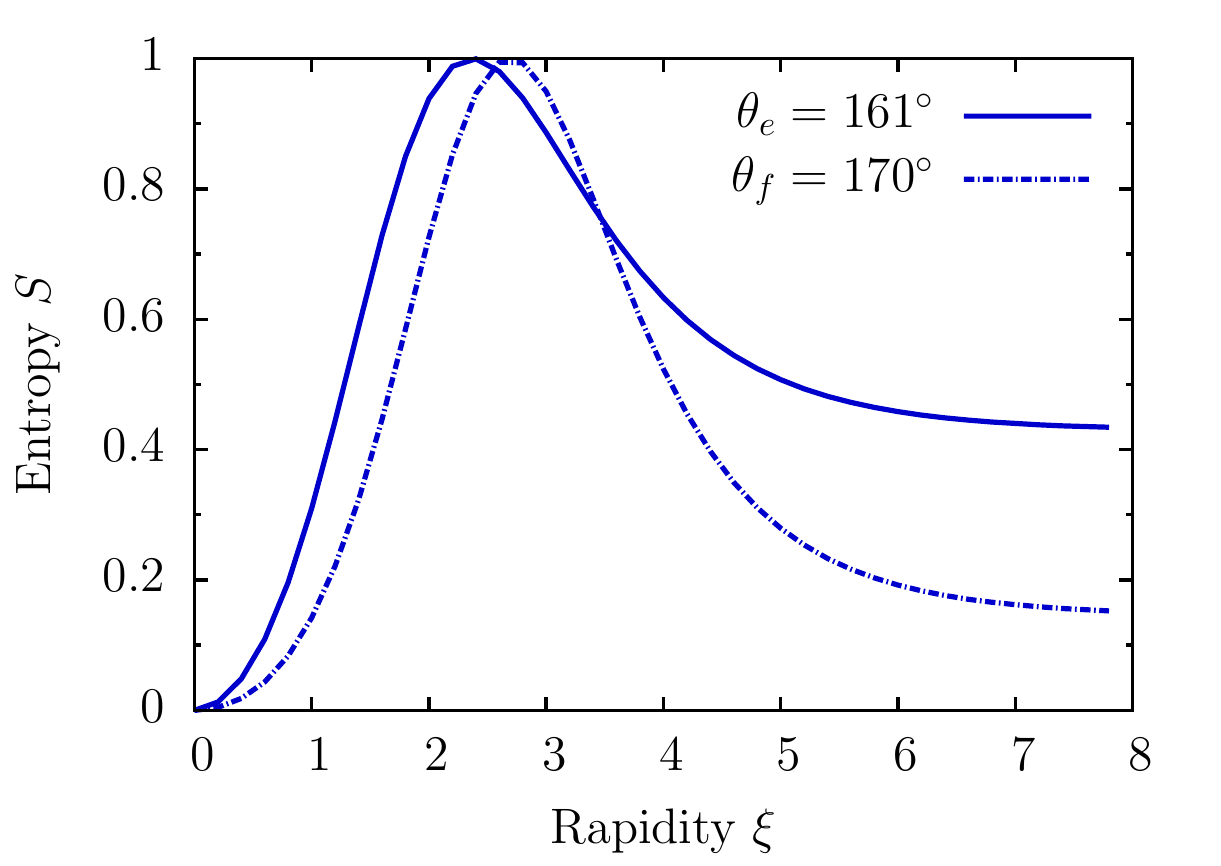}\label{fig:fig_2b_two_boosts}}%
\hspace{0.5em}
\subfigure[(Color online) Boost angles $\theta_a < 90\degree$, $\theta_b = 90\degree$ and $\theta_a > 90\degree$ correspond to rest frame momenta $\pp_0$ and are are shown for one peak of each state.]{\includegraphics[width=0.25\textwidth]{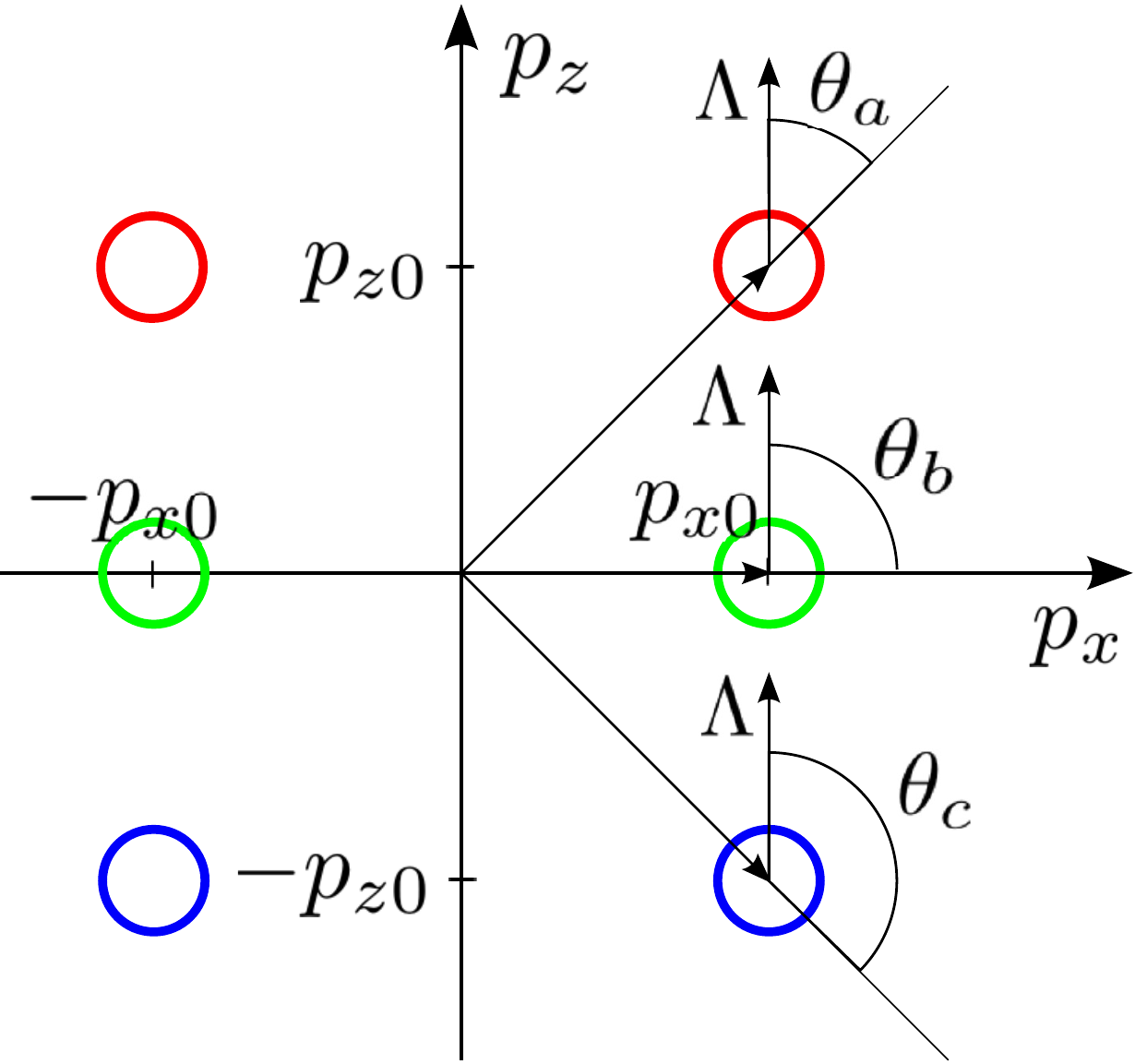}%
\label{fig:fig_2c_geometry}}%
\caption{(Color online) (a) and (b) Spin entropy for $x$-symmetric Gaussians with $\sigma / m = 1$. (c) Schematic representation of Gaussians in the rest frame, centered at different $\pp_0 = (\pm p_{x0}, 0, p_{z0})$ in the momentum space. Boost $\Lambda \equiv \Lambda(\xi)$ is in the positive $z$-direction. The width of the Gaussians shown is not to scale.}%
\label{fig:fig_2_twoPeakGaussian}
\end{figure*}
\end{widetext}

In the first, the particle has a momentum component $p_{z0}$ in the same direction as the boost. Thus the centers of Gaussians $\pp_0 = (\pm p_{x0}, 0, p_{z0})$ make angles of $\theta_a < 90\degree$ to the direction of boost, see FIG.~\ref{fig:fig_2c_geometry}. In the second scenario, the initial momenta $\pp_0 = (\pm p_{x0}, 0, 0)$ are orthogonal to the boost, 
%
%
%
%
so $\theta_b = 90\degree$. In the third, the particle's momentum has a $p_{z0}$ component opposite to the boost direction, hence $\theta_c > 90\degree$ and $\pp_0 = (\pm p_{x0}, 0, -p_{z0})$. In order to see how much entanglement has changed between spin and momentum, we 
plot spin entropy $S(\rho_S)$ for all scenarios in FIG.~\ref{fig:fig_2_twoPeakGaussian}.

These exhibit interesting properties.
Whereas previously \cite{dunningham_entanglement_2009} we found that spin entropy increases with boosts and attains the maximum value 1 as $v_2 \rightarrow 1$, 
results here confirm the above hypothesis that change of entanglement is sensitive to the direction and magnitude of boost. 
A general feature present in all scenarios is that spin-momentum entanglement initially increases with both boosts and later on saturates at a particular level.  
It is intriguing however that in some geometries maximal entanglement can be reached before the speed of light (FIG.~\ref{fig:fig_2a_three_boosts} and \ref{fig:fig_2b_two_boosts}), viz.\ when boost angle $\theta \geq 90\degree$ and surprisingly, that further increase of boost angle and magnitude may cause significant deterioration of entanglement (FIG.~\ref{fig:fig_2b_two_boosts}).

\begin{figure}[htb]
\includegraphics[width=0.25\textwidth]{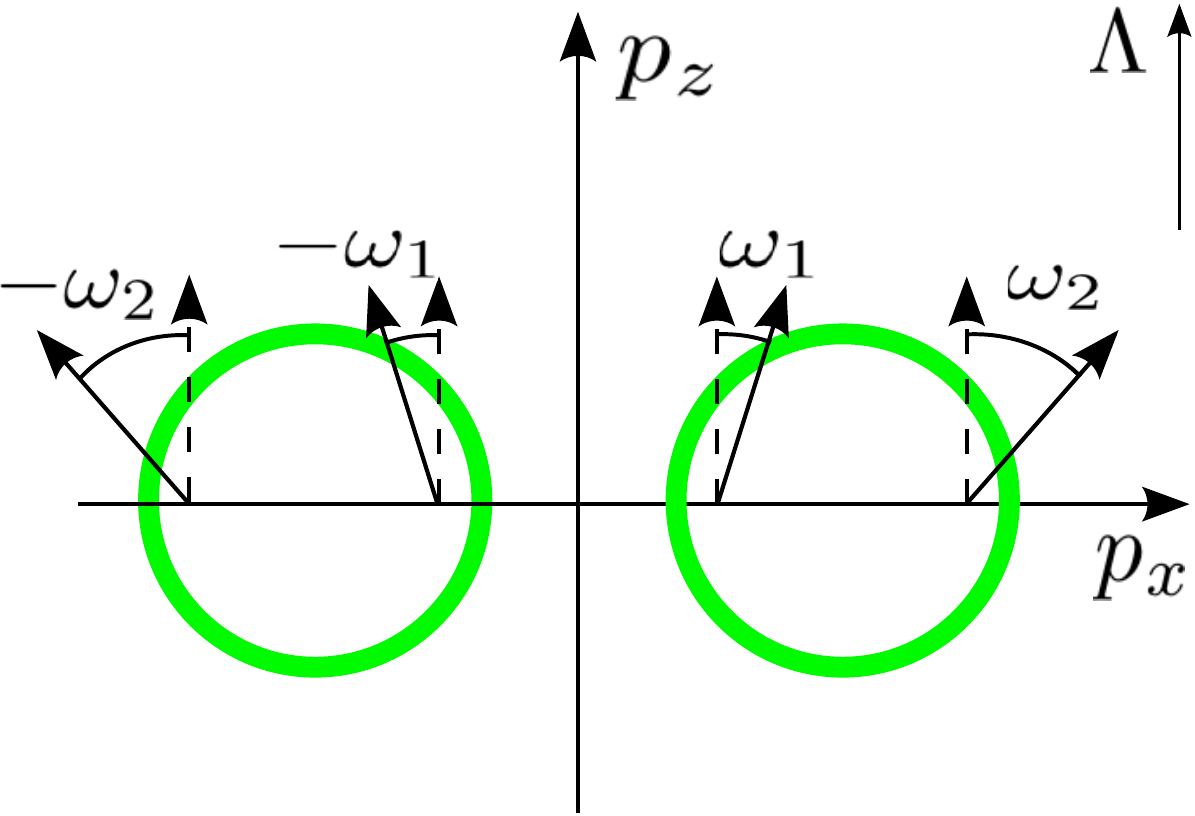}
\caption{\label{fig:fig_3_double_spin_field}(Color online) In the rest frame, the Gaussian spin field (circle) is given by a constant field of $z$-up spins (dashed). In the boosted frame, each spin $\lambda(\pp_i)$ of the field is Wigner rotated by a particular $\omega_i \equiv \omega(\pp_i)$. For a fixed boost $\xi$, 
rotation angle increases with $|\pp_i|$. Boost $\Lambda \equiv \Lambda(\xi)$ is in the positive $z$-direction.
}
\end{figure}

\emph{Spin and momentum from a geometric point of view}\label{sec:SpinAndMomentumFromAGeometricPointOfView}.---%
To understand the behavior of entanglement, it is useful to adopt a geometric perspective. One can think of vectors $\ket{\pp}\ket{\lambda}$ in Hilbert space as vector fields $\lambda(\pp)$ on the mass-shell of a particle with mass $m$. 
Whereas the geometric picture applies to both the continuous and discrete case, the essential qualitative behavior can be understood in terms of a discrete model of four spins in FIG.~\ref{fig:fig_3_double_spin_field} which we will use from now on. 
The spin state $\rho_S$, found by tracing out momentum, can be can be viewed as taking a (possibly infinite) convex sum of spin projection operators $\ket{\lambda(\pp)}\bra{\lambda(\pp)} = \Pi_{\lambda}(\pp)$ over the support of the Gaussian.
In our discrete example this reduces to 
\begin{align}\label{discreteSumFourSpinProjectors}
\rho_S = \alpha(-\pp_2) &\Pi_{\lambda}(-\pp_2) + \alpha(-\pp_1) \Pi_{\lambda}(-\pp_1) \nonumber\\ 
&+ \alpha(\pp_1) \Pi_{\lambda}(\pp_1) + \alpha(\pp_2) \Pi_{\lambda}(\pp_2) \;,
\end{align}
where the coefficients satisfy $\sum_{i} \alpha(\pp_i) = 1.$

\begin{figure}[htb]
\includegraphics[width=0.15\textwidth]{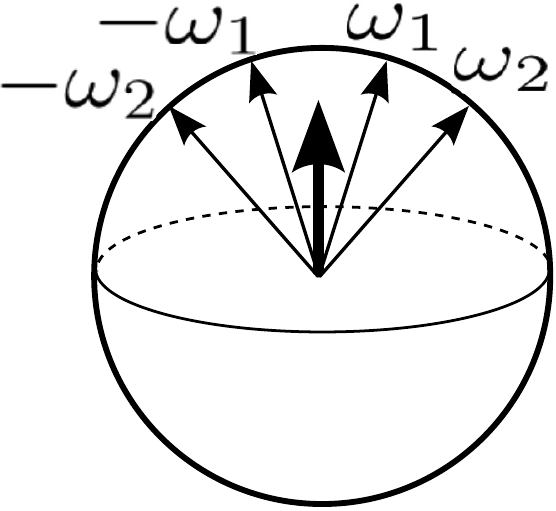}
\caption{\label{fig:fig_4_spin_on_bloch}
Tracing out momentum amounts to forming a convex sum of spins $\Pi_{\lambda}(\pp_i)$ that are Wigner rotated by $\omega_i \equiv \omega(\pp_i)$, here represented on the Bloch sphere. The resulting spin state $\rho_S$ (boldface arrow) is generally mixed.   
}
\end{figure}

It is now relatively easy to see how entanglement between spin and momentum arises.
Suppose the rest frame state is given by a product of spin and momentum as in Eq.\ (\ref{eq:zSpinUpGaussianMomentum}). 
This corresponds to a constant spin (operator) field in the momentum space, depicted by dashed arrows in FIG.~\ref{fig:fig_3_double_spin_field}. When the field is Lorentz boosted, each individual spin $\lambda(\pp)$ in FIG.~\ref{fig:fig_3_double_spin_field} is rotated by a different Wigner angle $\omega_i$, whose magnitude is determined by $|\pp_i|$, boost $\xi$ and the angle $\theta$ between $\pp$ and the boost direction. 
Hence after the boost each spin in the momentum space points in a different direction and the total state does not factorize any more: spin and momentum have become entangled. 
This means the spin operators $\Pi_{\lambda}(\pp_i)$ on the Bloch sphere in FIG.~\ref{fig:fig_4_spin_on_bloch} also point to different directions and summing them up yields in general a mixed state $\rho_S$.
Combined with the properties of Wigner rotation, we can now explain all the qualitative features of spin-momentum entanglement in FIG.~\ref{fig:fig_2a_three_boosts} and \ref{fig:fig_2b_two_boosts}---saturation, its level, and whether or not there is a bump.

\emph{Saturation.}---%
Saturation was first noted in \citep{gingrich_quantum_2002} where the authors study two spin-$1/2$ particles in a Bell state with a Gaussian product momentum as an initial state. Our results confirm that saturation occurs for a single particle with a Gaussian product momentum. 
The reason can be traced back to the properties of Wigner rotation. Given any two boosts at a particular angle $\theta$, when both boosts approach the speed of light, Wigner rotation asymptotically approaches a particular maximum value $\omega_m$ (see, for example, FIG.~\ref{fig:fig_1_wigner_rotation}). This implies that each individual spin of the field asymptotically approaches a particular $\pp$-dependent maximum rotation angle $\omega_{m}(\pp)$ as both boosts 
approach the speed of light. Since entropy is a monotonic function of spin, its behavior follows the same pattern: entropy approaches asymptotically a particular level as rapidity grows arbitrarily large.

\emph{Level of saturation.}---%
Although this explains why saturation occurs, it requires some qualification to account for why saturation reaches \emph{different levels} for Gaussians initially centered at different $p_{z0}$. 
This originates in the fact that the maximum value of Wigner rotation $\omega_m$ depends on the angle $\theta$ between two boosts. In our boosting scheme, the second boost is always in the $z$-direction. This means boost angle $\theta$ is determined by the center $\pp_0$ of the Gaussian wave packet. 
However, specifying $\theta$ amounts to setting a bound on the maximum value of rotation, that is, specifying $\omega_m$. The latter, in turn, sets a bound to the maximum rotation of spin operators on the Bloch sphere in FIG.~\ref{fig:fig_4_spin_on_bloch} or, equivalently, entropy. 
As a result, for two Gaussians with angles $\theta_a$ and $\theta_b$, where $\theta_a < \theta_b$, entanglement saturates at a lower level for $\theta_a$ than for $\theta_b$.

\emph{The bump effect.}---%
For boost geometries with $\theta \geq 90\degree$ entanglement initially reaches a maximum value and thereafter saturates at a lower value. It might seem that this contradicts what we just said about saturation. In light of the spin field picture, however, the bump is to be expected in such boost geometries. By way of example, consider the scenario with $v_1 = 0.999$, $\theta = 161\degree$ in FIG.~\ref{fig:fig_2b_two_boosts}.
Initially, as rapidity starts to grow, spins start to rotate in opposite directions at either Gaussian and so entanglement starts to increase in line with the explanation above.  
At $\xi = 2.4$, the effective spin of either Gaussian in FIG.~\ref{fig:fig_4_spin_on_bloch} has rotated by $|\omega| = 90\degree$, hence the spins of the left and right Gaussians become orthogonal and entanglement attains the maximum value 1. Now as rapidity increases further, spins `over-rotate', becoming again non-orthogonal and spin entropy starts to decrease. 
As rapidity grows even larger, the Wigner rotation attains a maximum value $\omega_m$ and entropy saturates at a value less than 1. The larger the $\theta$, the larger is $\omega_m$ and the lower is the final level of saturation as is seen in FIG.~\ref{fig:fig_2b_two_boosts}.
In the limiting case of large boosts $v_1, v_2 \rightarrow 1$, narrow Gaussians, $\sigma \rightarrow 0$ and boost angles $\theta \rightarrow 180\degree$, the boosted state approaches a product state and entanglement vanishes.

\emph{Conclusions.}---%
We establish that maximal entanglement between spin and momentum components of a single particle can be achieved with sub-luminal boosts. However, due to rich geometric setting, boost parameters must be chosen carefully as too large boosts lead to deterioration of entanglement. The effect persists for realistic states, i.e.\ Gaussian wave packets. Furthermore, all the diverse qualitative features of entanglement behavior can be given a natural geometric explanation, which could also be extended to an analysis of multiparticle entanglement.

\emph{Acknowledgments.}---%
This work was financially supported by the United Kingdom EPSRC.

\bibliography{SINGLE_particle_entanglement}

\begin{thebibliography}{12}%
\makeatletter
\providecommand \@ifxundefined [1]{%
 \@ifx{#1\undefined}
}%
\providecommand \@ifnum [1]{%
 \ifnum #1\expandafter \@firstoftwo
 \else \expandafter \@secondoftwo
 \fi
}%
\providecommand \@ifx [1]{%
 \ifx #1\expandafter \@firstoftwo
 \else \expandafter \@secondoftwo
 \fi
}%
\providecommand \natexlab [1]{#1}%
\providecommand \enquote  [1]{``#1''}%
\providecommand \bibnamefont  [1]{#1}%
\providecommand \bibfnamefont [1]{#1}%
\providecommand \citenamefont [1]{#1}%
\providecommand \href@noop [0]{\@secondoftwo}%
\providecommand \href [0]{\begingroup \@sanitize@url \@href}%
\providecommand \@href[1]{\@@startlink{#1}\@@href}%
\providecommand \@@href[1]{\endgroup#1\@@endlink}%
\providecommand \@sanitize@url [0]{\catcode `\\12\catcode `\$12\catcode
  `\&12\catcode `\#12\catcode `\^12\catcode `\_12\catcode `\%12\relax}%
\providecommand \@@startlink[1]{}%
\providecommand \@@endlink[0]{}%
\providecommand \url  [0]{\begingroup\@sanitize@url \@url }%
\providecommand \@url [1]{\endgroup\@href {#1}{\urlprefix }}%
\providecommand \urlprefix  [0]{URL }%
\providecommand \Eprint [0]{\href }%
\providecommand \doibase [0]{http://dx.doi.org/}%
\providecommand \selectlanguage [0]{\@gobble}%
\providecommand \bibinfo  [0]{\@secondoftwo}%
\providecommand \bibfield  [0]{\@secondoftwo}%
\providecommand \translation [1]{[#1]}%
\providecommand \BibitemOpen [0]{}%
\providecommand \bibitemStop [0]{}%
\providecommand \bibitemNoStop [0]{.\EOS\space}%
\providecommand \EOS [0]{\spacefactor3000\relax}%
\providecommand \BibitemShut  [1]{\csname bibitem#1\endcsname}%
\let\auto@bib@innerbib\@empty
\bibitem [{\citenamefont {Dunningham}\ \emph {et~al.}(2009)\citenamefont
  {Dunningham}, \citenamefont {Palge},\ and\ \citenamefont
  {Vedral}}]{dunningham_entanglement_2009}%
  \BibitemOpen
  \bibfield  {author} {\bibinfo {author} {\bibfnamefont {J.}~\bibnamefont
  {Dunningham}}, \bibinfo {author} {\bibfnamefont {V.}~\bibnamefont {Palge}}, \
  and\ \bibinfo {author} {\bibfnamefont {V.}~\bibnamefont {Vedral}},\
  }\href@noop {} {\bibfield  {journal} {\bibinfo  {journal} {Phys. Rev. A}\
  }\textbf {\bibinfo {volume} {80}},\ \bibinfo {pages} {044302} (\bibinfo
  {year} {2009})}\BibitemShut {NoStop}%
\bibitem [{\citenamefont {Peres}\ \emph {et~al.}(2002)\citenamefont {Peres},
  \citenamefont {Scudo},\ and\ \citenamefont {Terno}}]{peres_quantum_2002}%
  \BibitemOpen
  \bibfield  {author} {\bibinfo {author} {\bibfnamefont {A.}~\bibnamefont
  {Peres}}, \bibinfo {author} {\bibfnamefont {P.~F.}\ \bibnamefont {Scudo}}, \
  and\ \bibinfo {author} {\bibfnamefont {D.~R.}\ \bibnamefont {Terno}},\
  }\href@noop {} {\bibfield  {journal} {\bibinfo  {journal} {Phys. Rev. Lett.}\
  }\textbf {\bibinfo {volume} {88}} (\bibinfo {year} {2002})}\BibitemShut
  {NoStop}%
\bibitem [{\citenamefont {Gingrich}\ and\ \citenamefont
  {Adami}(2002)}]{gingrich_quantum_2002}%
  \BibitemOpen
  \bibfield  {author} {\bibinfo {author} {\bibfnamefont {R.~M.}\ \bibnamefont
  {Gingrich}}\ and\ \bibinfo {author} {\bibfnamefont {C.}~\bibnamefont
  {Adami}},\ }\href@noop {} {\bibfield  {journal} {\bibinfo  {journal} {Phys.
  Rev. Lett.}\ }\textbf {\bibinfo {volume} {89}} (\bibinfo {year}
  {2002})}\BibitemShut {NoStop}%
\bibitem [{\citenamefont {Alsing}\ and\ \citenamefont
  {Milburn}(2002)}]{alsing_entanglement_2002}%
  \BibitemOpen
  \bibfield  {author} {\bibinfo {author} {\bibfnamefont {P.~M.}\ \bibnamefont
  {Alsing}}\ and\ \bibinfo {author} {\bibfnamefont {G.~J.}\ \bibnamefont
  {Milburn}},\ }\href@noop {} {\bibfield  {journal} {\bibinfo  {journal}
  {Quant. Inf. and Comp.}\ }\textbf {\bibinfo {volume} {2}},\ \bibinfo {pages}
  {487} (\bibinfo {year} {2002})}\BibitemShut {NoStop}%
\bibitem [{\citenamefont {Li}\ and\ \citenamefont
  {Du}(2003)}]{li_relativistic_2003}%
  \BibitemOpen
  \bibfield  {author} {\bibinfo {author} {\bibfnamefont {H.}~\bibnamefont
  {Li}}\ and\ \bibinfo {author} {\bibfnamefont {J.}~\bibnamefont {Du}},\
  }\href@noop {} {\bibfield  {journal} {\bibinfo  {journal} {Phys. Rev. A}\
  }\textbf {\bibinfo {volume} {68}} (\bibinfo {year} {2003})}\BibitemShut
  {NoStop}%
\bibitem [{\citenamefont {Peres}\ and\ \citenamefont
  {Terno}(2004)}]{peres_quantum_2004}%
  \BibitemOpen
  \bibfield  {author} {\bibinfo {author} {\bibfnamefont {A.}~\bibnamefont
  {Peres}}\ and\ \bibinfo {author} {\bibfnamefont {D.}~\bibnamefont {Terno}},\
  }\href@noop {} {\bibfield  {journal} {\bibinfo  {journal} {Rev. Mod. Phys.}\
  }\textbf {\bibinfo {volume} {76}},\ \bibinfo {pages} {93} (\bibinfo {year}
  {2004})}\BibitemShut {NoStop}%
\bibitem [{\citenamefont {Caban}\ and\ \citenamefont
  {Rembieli{\'n}ski}(2005)}]{caban_lorentz-covariant_2005}%
  \BibitemOpen
  \bibfield  {author} {\bibinfo {author} {\bibfnamefont {P.}~\bibnamefont
  {Caban}}\ and\ \bibinfo {author} {\bibfnamefont {J.}~\bibnamefont
  {Rembieli{\'n}ski}},\ }\href@noop {} {\bibfield  {journal} {\bibinfo
  {journal} {Physical Review A}\ }\textbf {\bibinfo {volume} {72}},\ \bibinfo
  {pages} {012103} (\bibinfo {year} {2005})}\BibitemShut {NoStop}%
\bibitem [{\citenamefont {Caban}\ and\ \citenamefont
  {Rembieli{\'n}ski}(2006)}]{caban_einstein-podolsky-rosen_2006-1}%
  \BibitemOpen
  \bibfield  {author} {\bibinfo {author} {\bibfnamefont {P.}~\bibnamefont
  {Caban}}\ and\ \bibinfo {author} {\bibfnamefont {J.}~\bibnamefont
  {Rembieli{\'n}ski}},\ }\href@noop {} {\bibfield  {journal} {\bibinfo
  {journal} {Physical Review A}\ }\textbf {\bibinfo {volume} {74}},\ \bibinfo
  {pages} {042103} (\bibinfo {year} {2006})}\BibitemShut {NoStop}%
\bibitem [{\citenamefont {Jordan}\ \emph {et~al.}(2007)\citenamefont {Jordan},
  \citenamefont {Shaji},\ and\ \citenamefont
  {Sudarshan}}]{jordan_lorentz_2007}%
  \BibitemOpen
  \bibfield  {author} {\bibinfo {author} {\bibfnamefont {T.}~\bibnamefont
  {Jordan}}, \bibinfo {author} {\bibfnamefont {A.}~\bibnamefont {Shaji}}, \
  and\ \bibinfo {author} {\bibfnamefont {E.}~\bibnamefont {Sudarshan}},\
  }\href@noop {} {\bibfield  {journal} {\bibinfo  {journal} {Physical Review
  A}\ }\textbf {\bibinfo {volume} {75}} (\bibinfo {year} {2007})}\BibitemShut
  {NoStop}%
\bibitem [{\citenamefont {Friis}\ \emph {et~al.}(2010)\citenamefont {Friis},
  \citenamefont {Bertlmann},\ and\ \citenamefont
  {Huber}}]{friis_relativistic_2010}%
  \BibitemOpen
  \bibfield  {author} {\bibinfo {author} {\bibfnamefont {N.}~\bibnamefont
  {Friis}}, \bibinfo {author} {\bibfnamefont {R.~A.}\ \bibnamefont
  {Bertlmann}}, \ and\ \bibinfo {author} {\bibfnamefont {M.}~\bibnamefont
  {Huber}},\ }\href@noop {} {\bibfield  {journal} {\bibinfo  {journal}
  {Physical Review A}\ }\textbf {\bibinfo {volume} {81}} (\bibinfo {year}
  {2010})}\BibitemShut {NoStop}%
\bibitem [{\citenamefont {Rhodes}\ and\ \citenamefont
  {Semon}(2004)}]{rhodes_relativistic_2004}%
  \BibitemOpen
  \bibfield  {author} {\bibinfo {author} {\bibfnamefont {J.~A.}\ \bibnamefont
  {Rhodes}}\ and\ \bibinfo {author} {\bibfnamefont {M.~D.}\ \bibnamefont
  {Semon}},\ }\href@noop {} {\bibfield  {journal} {\bibinfo  {journal} {Am. J.
  Phys.}\ }\textbf {\bibinfo {volume} {72}},\ \bibinfo {pages} {943} (\bibinfo
  {year} {2004})}\BibitemShut {NoStop}%
\bibitem [{\citenamefont {Halpern}(1968)}]{halpern_special_1968}%
  \BibitemOpen
  \bibfield  {author} {\bibinfo {author} {\bibfnamefont {F.~R.}\ \bibnamefont
  {Halpern}},\ }\href@noop {} {\emph {\bibinfo {title} {Special Relativity and
  Quantum Mechanics}}}\ (\bibinfo  {publisher} {{Prentice-Hall}},\ \bibinfo
  {year} {1968})\BibitemShut {NoStop}%
\end{thebibliography}%

\end{document}